\documentclass[12pt]{iopart}
\usepackage{epsfig}
\usepackage{color}
\usepackage{graphics}
\usepackage{iopams}
\usepackage{mathrsfs}

\begin{document}

\title[Generalized diffusion-wave equation with memory kernel]{Generalized diffusion-wave equation with memory kernel}

\author{Trifce Sandev$^{1,2,3}$, Zivorad Tomovski$^4$, Johan L.A. Dubbeldam$^5$, Aleksei Chechkin$^{6,7}$}

\address{$^{1}$Radiation Safety Directorate, Partizanski odredi 143, P.O. Box 22, 1020 Skopje, Macedonia\\ $^{2}$Institute of Physics, Faculty of Natural Sciences and Mathematics, Ss Cyril and Methodius University, Arhimedova 3, 1000 Skopje, Macedonia\\ $^{3}$Research Center for Computer Science and Information Technologies, Macedonian Academy of Sciences and Arts, Bul. Krste Misirkov 2, 1000 Skopje, Macedonia\\ $^4$Faculty of Natural Sciences and Mathematics, Institute of Mathematics, Saints Cyril and Methodius University, 1000 Skopje, Macedonia\\ $^5$Delft Institute of Applied Mathematics, Delft Universiy of Technology, Mekelweg 4, 2628CD Delft, The Netherlands\\ $^{6}$Akhiezer Institute for Theoretical Physics, National Science Center ``Kharkov Institute of Physics and Technology", Kharkov 61108, Ukraine\\ $^{7}$Institute for Physics and Astronomy, University of Potsdam, D-14776 Potsdam-Golm, Germany}

\ead{trifce.sandev@drs.gov.mk}

\begin{abstract}
We study generalized diffusion-wave equation in which the second order time derivative is replaced by integro-differential operator. It yields time fractional and distributed order time fractional diffusion-wave equations as particular cases. We consider different memory kernels of the integro-differential operator, derive corresponding fundamental solutions, specify the conditions of their non-negativity and calculate mean squared displacement for all cases. In particular, we introduce and study generalized diffusion-wave equations with regularized Prabhakar derivative of single and distributed orders. The equations considered can be used for modeling broad spectrum of anomalous diffusion processes and various transitions between different diffusion regimes.
\end{abstract}

\maketitle

\section{Introduction}

Diffusion equations with fractional time and space derivatives instead of the integer ones are widely used to describe anomalous diffusion processes where the mean squared displacement (MSD) scales as a power of time,
\begin{eqnarray}\label{msd intro}
\left\langle x^{2}(t)\right\rangle\simeq t^{\alpha},
\end{eqnarray}
where $\alpha\neq 1$ \cite{Metzler3}. If $0 < \alpha < 1$, the process is subdiffusive, if $1 < \alpha < 2$ the process is superdiffusive. Ballistic diffusion corresponds to $\alpha = 2$, whereas $\alpha > 2$ corresponds to superballlistic motion. At $\alpha = 1$ one has normal Brownian diffusion that obeys the usual diffusion equation with the time and space derivatives of the first and second orders, respectively. Classical examples of subdiffusion include charge carrier motion in semiconductors \cite{scher}, spreading of tracer chemicals in subsurface aquifers \cite{berkowitz}, as well as the motion of passive particles in atmospheric convection rolls \cite{young}. Superdiffusion is known from tracer motion in chaotic laminar flows, which takes place due to vortices working as traps \cite{solomon}, plasma turbulence in fusion devices \cite{negrete}, diffusion in porous structurally inhomogeneous media \cite{new}, or from random search, like actively moving bacteria \cite{ariel} and animals \cite{viswanathan}, or from human travels \cite{brockmann}.

Modern microscopic techniques such as fluorescence correlation spectroscopy or advanced single particle tracking methods have led to the discovery of a multitude of anomalous diffusion processes in living biological cells and complex fluids, see e.g., the reviews \cite{barkai,franosch,meroz,PCCP,lene,saxton} and references therein. With the number of anomalous diffusion phenomena growing it became clear that most of the complex systems do not show a mono-scaling behavior, Eq.(\ref{msd intro}), but instead demonstrate transitions between different diffusion regimes in course of time. Such observations put forward the idea that in order to capture such multi-scaling situations one could replace relatively simple operators of fractional derivatives by more general operators with specific memory kernels. Then, a special case of a power-law kernel recovers fractional derivative and respectively, the mono-scaling diffusion regime. Thus, the distributed order fractional diffusion equations in the normal and modified forms have been suggested in order to describe the processes which become more anomalous (retarding subdiffusion or accelerating superdiffusion) or less anomalous (accelerating subdiffusion and decelerating superdiffusion) in course of time \cite{chechkin2,chechkin,chechkin3,chechkin4,chechkin5,sokolov}. The solutions of such distributed order equations have been obtained, and the relation to continuous time random walk models has been investigated in detail \cite{gorenflo,GoMa,MaPa,MaPaGo,fcaa2015,fcaa2018,pre2015}. Other forms of kernels contain for example, distributed, tempered, distributed tempered and Mittag-Leffler memory functions \cite{chechkin5,kochubei,fcaa2015,fcaa2018,csf2017,stan}, to name but a few.

In analogy to the generalized diffusion equation with memory kernel introduced in \cite{fcaa2015} (see also \cite{csf2017}), here we introduce the following {\it generalized diffusion-wave equation}
\begin{eqnarray}\label{distributed order wave eq memory}
\int_{0}^{t}\eta(t-t')\frac{\partial^{2}}{\partial t'^{2}}W(x,t)\,dt'=\frac{\partial^{2}}{\partial x^{2}}W(x,t),
\end{eqnarray}
with non-negative memory kernel $\eta(t)$ of physical dimension $[\eta(t)]=\mathrm{m}^{-2}\mathrm{s}^{1}$, where $W(x,t)$ is the field variable. In what follows for simplicity we use dimensionless units. Here we note that $W(x,t)$ could be either a probability distribution function (PDF) of particles which should be non-negative, or electric field or velocity, where the non-negativity might not be needed. In Ref.~\cite{fcaa2015} a similar equation -- called {\it generalized diffusion equation} -- was suggested, that contains a first order time derivative in the integrand, and the corresponding continuous time random walk model was also developed. Different generic forms of the memory kernel have been employed (power-law, truncated power-law, Mittag-Leffler and truncated Mittag-Leffler) in order to demonstrate the variety of anomalous diffusion regimes that could be described within the suggested generalized diffusion equation. The present paper can be considered as a natural continuation and extension of such kind of research aiming to give a comprehensive description of anomalous diffusion phenomena. In what follows we thus consider $W(x,t)$ as a PDF and study solutions of Eq.~(\ref{distributed order wave eq memory}) in the infinite domain with zero boundary conditions at infinity, $W(\pm\infty,t)=0$, $\frac{\partial}{\partial x}W(\pm\infty,t)=0$. We take the initial conditions of the form
\begin{eqnarray}\label{initial_condition}
W(x,t=0)= \delta(x), \quad \frac{\partial}{\partial t}W(x,t=0)= 0.
\end{eqnarray}
The first initial condition is obvious since it corresponds to the particle initially starting at $x=0$. This is a standard set-up used in the papers on anomalous diffusion processes where the mean square displacement is calculated as a function of time in the infinite space domain. Second initial condition must be imposed on $\frac{\partial}{\partial t}W(x,t)$ because of the second time derivative that appears in the generalized diffusion wave equation (\ref{distributed order wave eq memory}) in the integrand of the left hand side. To the authors opinion the concretization of such condition is not a trivial task. Similar problem appeared for telegrapher's equation \cite{masoliver_fin,weiss} and its fractional generalizations reproducing anomalous diffusion laws \cite{metz1,metz2,metz3}. In Refs.~\cite{masoliver_fin,weiss} the zero second initial condition was derived from the underlining random walk scheme under the specific assumption that the initial tendency to move in one direction or another is non-biased. In Refs.~\cite{metz1,metz2,metz3} the zero second initial condition was also employed implicitly without addressing to the corresponding continuous time random walk model. To the authors knowledge at present time a stochastic approach to the generalized diffusion-wave equation (\ref{distributed order wave eq memory}) is lacking (note that instead, for the generalized diffusion equation such approach was recently developed in Refs.~\cite{fcaa2015,fcaa2018}). Under these circumstances, in what follows we pay main attention to the particular case of the zero second initial condition, Eq.~(\ref{initial_condition}). However, in Section 4 we provide discussion of how the anomalous diffusion regimes found in the paper can be modified in presence of non-zero second initial condition, and which restrictions on such initial condition must be imposed. We note that for the special case of the memory kernel $$\eta(t)=\int_{1}^{2}p(\lambda)\frac{t^{1-\lambda}}{\Gamma(2-\lambda)}\,d\lambda,$$ where the weight function $p(\lambda)$ satisfies $\int_{1}^{2}p(\lambda)\,d\lambda=1$, one recovers the distributed order wave equation considered by Gorenflo, Luchko and Stojanovic \cite{gorenflo fcaa2013}, i.e.,
\begin{eqnarray}\label{distributed order wave eq}
\int_{1}^{2}p(\lambda)\,{_{C}D_{t}^{\lambda}}W(x,t)\,d\lambda=\frac{\partial^{2}}{\partial x^{2}}W(x,t), \quad 1<\lambda<2,
\end{eqnarray}
where ${_{C}D_{t}^{\lambda}}$ is the Caputo fractional derivative of order $\lambda$ \cite{podlubny}
\begin{eqnarray}\label{Caputo_derivative}
{_{C}D_{t}^{\lambda}}f(t)=\frac{1}{\Gamma(n-\lambda)}\int_{0}^{t}(t-t')^{n-1-\lambda}\frac{d^{n}}{dt'^{n}}f(t')\,dt',
\end{eqnarray}
$n-1<\lambda<n$, $n\in N$. The case with $p(\lambda)=\delta(\lambda-\mu)$ recovers fractional diffusion-wave equation which was studied in different contexts in \cite{gorenflo fcaa2013,meerschaert,meerschaert2,mainardi,luchko mainardi,bazhlekova,bazhlekova2,schneider}. In \cite{gorenflo fcaa2013} the authors derived the fundamental solutions of Eq.~(\ref{distributed order wave eq}) for different forms of the weight function, and discussed the non-negativity of the solution by using the properties of completely monotone, Bernstein and Stieltjes functions. In this paper we introduce other generalized memory kernels and analyse properties of the solutions. 

The paper is organized as follows. In Section 2 we derive the solution of the generalized diffusion-wave equation (\ref{distributed order wave eq memory}). We find the constraints on the memory kernel, which guarantee non-negativity of the solution. We give general formula for the moments of the fundamental solution. Special cases of the memory kernel are considered in Section 3, and the corresponding solutions and second moments are obtained. In Section 4 we discuss the role of a second non-zero initial condition. The summary is given in Section 5.

\section{General solution and its fractional moments}

\subsection{Completely monotone and Bernstein functions}

Before finding fundamental solution of the model proposed we recall definitions and some properties of the completely monotone (CM) and Bernstein functions (BF) \cite{book bernstein}, which are used to show the non-negativity of the fundamental solution.

\subsubsection{Completely monotone functions.}

Completely monotone function $\mathcal{M}(s)$ is defined in non-negative semi-axis and have a property that 
\begin{equation}\label{definition CM}
(-1)^n \mathcal{M}^{(n)}(s) \geq 0 \quad \textrm{for all $n\in\mathrm{N}_{0}$ and $s\geq 0$}.
\end{equation}

According to the famous Bernstein theorem the completely monotone function can be represented as a Laplace transform of non-negative function $p(t)$ (see, for example, Chapter XIII, Section 4 in Ref.~\cite{feller}), i.e., 
\begin{equation}\label{CM NN}
\mathcal{M}(s) = \int_0^\infty p(t)e^{-st}\,dt.
\end{equation} 

The following properties hold true for CM functions:
\begin{itemize}
 \item[({\bf a})] Linear combination $a_1 \mathcal{M}_{1}(s)+a_2 \mathcal{M}_{2}(s)$ ($a_{1},a_{2}\geq0$) of completely monotone functions $\mathcal{M}_{1}(s)$ and $\mathcal{M}_{2}(s)$ is completely monotone function as well;
 \item[({\bf b})] The product $\mathcal{M}(s)=\mathcal{M}_1(s)\mathcal{M}_2(s)$ of completely monotone functions $\mathcal{M}_1(s)$ and $\mathcal{M}_2(s)$ is again completely monotone function.
\end{itemize}
An example of completely monotone function is $s^{\alpha}$, where $\alpha\leq0$.

\subsubsection{Stieltjes functions.}

A function defined in the non-negative semi-axis, which is Laplace transform of CM function is called Stieltjes function (SF). The Stieltjes functions are a subclass of the completely monotone functions. For the SFs the following property holds true:
\begin{itemize}
 \item[({\bf c})] Linear combination $a_1 \mathcal{S}_{1}(s)+a_2 \mathcal{S}_{2}(s)$ ($a_{1},a_{2}\geq0$) of SFs $\mathcal{S}_{1}(s)$ and $\mathcal{S}_{2}(s)$ is SF;
% \item[d)] If $s(x)$ is a Stieltjes functions then the function $\frac{1}{s\left(\frac{1}{x}\right)}$ is Stieltjes function as well.
\end{itemize} 
An example of SF is $s^{\alpha-1}$, where $0\leq\alpha\leq1$.

\subsubsection{Bernstein functions.}

The Bernstein function is a non-negative function whose derivative is completely monotone, that is
$$(-1)^{(n-1)} \mathcal{B}^{(n)}(s) \geq 0 \quad \textrm{for all $n=1,2,\dots$}$$
BFs have the following properties:
\begin{itemize}
 \item[({\bf d})] Linear combination $a_1\mathcal{B}_1(s)+a_2\mathcal{B}_2(s)$ ($a_1,a_2 \geq 0$) of BFs $\mathcal{B}_1(s)$ and $\mathcal{B}_2(s)$ is a BF;
\item[({\bf e})] A composition $\mathcal{M}(\mathcal{B}(s))$ of CM function $\mathcal{M}(s)$ and BF $\mathcal{B}(s)$ is CM function.
\end{itemize}
From these properties it follows that the function $e^{-u\,\mathcal{B}(s)}$ is CM for $u>0$ if $\mathcal{B}(s)$ is a BF. 

An example of a BF is $s^{\alpha}$, where $0\leq\alpha\leq1$.

\subsubsection{Complete Bernstein functions.}
The complete Bernstein functions (CBFs) are subclass of BFs. A function $\mathcal{C}(s)$ on $(0,\infty)$ is a CBF if and only if $\mathcal{C}(s)/s$ is a SF. 

The following properties are valid for the complete Bernstein functions:
\begin{itemize}
\item[({\bf f})] A composition $\mathcal{S}(\mathcal{C}(s))$ of SF $\mathcal{S}(s)$ and CBF $\mathcal{C}(s)$ is a SF;
\item[({\bf g})] A composition $\mathcal{C}(\mathcal{S}(s))$ of CBF $\mathcal{C}(s)$ and SF $\mathcal{S}(s)$ is SF;
\item[({\bf h})] If $\mathcal{C}(s)$ is a CBF, then $\mathcal{C}(s)/s$ is SF.
\end{itemize}
An example of complete Bernstein function is $s^{\alpha}$, where $0<\alpha<1$.

\subsection{Fundamental solution}

At first let us consider the generalized diffusion-wave equation (\ref{distributed order wave eq memory}) with the memory kernel $\eta(t)$ of a general form. The restrictions on the memory kernel are discussed below. In order to find the solution of equation (\ref{distributed order wave eq memory}) we use the Fourier $(\tilde{F}(\kappa)=\int_{-\infty}^{\infty}f(x)e^{\imath\kappa x}\,dx)$ and Laplace $(\hat{f}(s)=\int_{0}^{\infty}f(t)e^{-st}\,dt)$ transformations. For initial conditions $W(x,0)=\delta(x)$ and $\frac{\partial}{\partial t}W(x,0)=0$, and zero boundary conditions we find
\begin{eqnarray}\label{distributed order wave eq Laplace-Fourier space M}
\tilde{\hat{W}}(\kappa,s)=\frac{s\hat{\eta}(s)}{s^{2}\hat{\eta}(s)+\kappa^{2}}.
\end{eqnarray}
By the inverse Fourier transformation of Eq.~(\ref{distributed order wave eq Laplace-Fourier space M}) we find the PDF in the Laplace space
\begin{eqnarray}\label{PDF L}
\hat{W}(x,s)=\frac{1}{2}\sqrt{\hat{\eta}(s)}\exp\left(-s\sqrt{\hat{\eta}(s)}|x|\right),
\end{eqnarray}
from which one can find either the exact form of $W(x,t)$ or (at least) its asymptotic behavior by using Tauberian theorems \cite{feller}. From Eq.~(\ref{PDF L}) one can easily conclude that the PDF is normalized to 1, i.e., $\int_{-\infty}^{\infty}W(x,t)\,dx=1$, since $\int_{-\infty}^{\infty}\hat{W}(x,s)\,dx=\frac{1}{s}$.

The non-negativity of the solution of the distributed order wave equation has been shown in \cite{gorenflo fcaa2013} by using the properties of CM functions and BFs. In order to find the constraints for the memory kernel $\eta(t)$ under which the considered generalized diffusion-wave equation (2) has a non-negative solution we use representation (\ref{PDF L}) in the Laplace space together with relation (\ref{CM NN}). The solution (\ref{PDF L}) can be considered as a product of two functions, $\frac{1}{2}\sqrt{\hat{\eta}(s)}$ and $\exp\left(-s\sqrt{\hat{\eta}(s)}|x|\right)$. In order to show non-negativity of the PDF $W(x,t)$ one should prove that its Laplace transform $\hat{W}(x,s)$ is a CM function. Thus, it is sufficient to prove that both functions $$\sqrt{\hat{\eta}(s)} \quad \mathrm{and} \quad \exp\left(-s\sqrt{\hat{\eta}(s)}|x|\right)$$ are CM (see property ({\bf b}) in Section 2.1.1.). Therefore, it is sufficient to show that $\sqrt{\hat{\eta}(s)}$ is CM function, and $s\sqrt{\hat{\eta}(s)}$ is BF, according to the property ({\bf e}) in Section 2.1.3. The non-negativity of the solution can also be shown by proving that the function $\sqrt{\hat{\eta}(s)}$ is SF, which is again a CM function, or that $s\sqrt{\hat{\eta}(s)}$ is CBF, which follows from the properties ({\bf f}) and ({\bf g}) in Section 2.1.4.

\smallskip

{\it {\bf Remark 1.}}
{\it Here we note that the subordination approach, which was used to prove the non-negativity of the PDF in case of fractional and distributed order diffusion equations \cite{gorenflo fcaa2013,fcaa2015}, can not be applied in case of generalized diffusion-wave equations. Indeed, the solution (\ref{distributed order wave eq Laplace-Fourier space M}) can be rewritten in a form of subordination integral as
\begin{eqnarray}
\tilde{\hat{W}}(k,s)=s\hat{\eta}(s)\int_{0}^{\infty}e^{-u[s^{2}\hat{\eta}(s)+k^{2}]}\,du=\int_{0}^{\infty}e^{-uk^{2}}\hat{G}(u,s)\,du,
\end{eqnarray}
where $$\hat{G}(u,s)=s\hat{\eta}(s)e^{-us^{2}\hat{\eta}(s)}.$$ Therefore, the PDF $W(x,t)$ becomes
\begin{eqnarray}
W(x,t)=\int_{0}^{\infty}\frac{e^{-\frac{x^{2}}{4u}}}{\sqrt{4\pi u}}G(u,t)\,du.
\end{eqnarray}
From here it follows that in order $W(x,t)$ to be non-negative, the function $G(u,t)$ should be non-negative, that is $\hat{G}(u,s)$ should be CM. For this it is sufficient to show that $s\hat{\eta}(s)$ is CM and $s^{2}\hat{\eta}(s)$ is a BF according to property $(\textrm{{\bf e}})$ in Section 2.1.3. Such approach works well in case of fractional and distributed order diffusion equations but, as we will see later, not in case of generalized diffusion-wave equation. This was already noted in \cite{gorenflo fcaa2013} while proving the non-negativity of the distributed order diffusion-wave equation that is a particular case of the generalized diffusion-wave equation.}

\smallskip

\subsection{Fractional moments}

From Eq.~(\ref{distributed order wave eq Laplace-Fourier space M}) we calculate the $q$-th moment $\left\langle
|x|^{q}(t)\right\rangle=\int_{-\infty}^{\infty}|x|^{q}W(x,t)\,dx=2\int_{0}^{\infty}x^{q}W(x,t)\,dx$ of the solution of Eq.~(\ref{distributed order wave eq memory}), which is given by
\begin{eqnarray}\label{moments}
\left\langle
|x|^{q}(t)\right\rangle=\Gamma(q+1)\mathcal{L}^{-1}\left[\frac{s^{-1}}{\left(s^{2}\hat{\eta}(s)\right)^{q/2}}\right](t).
\end{eqnarray}
For the second moment we find
\begin{eqnarray}\label{second moment}
\left\langle
x^{2}(t)\right\rangle=\left.\left\{-\frac{\partial^{2}}{\partial\kappa^{2}}\mathcal{L}^{-1}\left[\tilde{\hat{W}}(\kappa,s)\right](\kappa,t)\right\}\right|_{\kappa=0}=2\,\mathcal{L}^{-1}\left[\frac{1}{s^{3}\hat{\eta}(s)}\right](t),
\end{eqnarray}
from where one can analyze diffusion regimes for different forms of the memory function $\eta(t)$.

\section{Special cases}

In this Section we analyze solutions of the generalized diffusion-wave equation for different forms of the memory kernel. We also calculate the mean squared displacement for each of the cases.

\subsection{Standard wave equation. Dirac delta memory kernel}

Eq.~(\ref{distributed order wave eq}) for a Dirac delta memory kernel $\eta(t)=\delta(t)$ yields the classical wave equation
\begin{eqnarray}\label{classical cattaneo eq}
\frac{\partial^{2}}{\partial t^{2}}W(x,t)=\frac{\partial^{2}}{\partial x^{2}}W(x,t).
\end{eqnarray}
Its solution can be obtained from (\ref{distributed order wave eq Laplace-Fourier space M}) in the d'Alembert form,
\begin{eqnarray}\label{PDF L delta}
W(x,t)=\mathcal{F}^{-1}\left[\cos(t\,\kappa)\right]=\frac{1}{2}\left[\delta\left(x+t\right)+\delta\left(x-t\right)\right]. 
\end{eqnarray}
The non-negativity of the solution of standard wave equation (\ref{classical cattaneo eq}) is obvious since $\sqrt{\hat{\eta}(s)}=1$ is CM (see the example in Section 2.1.1), and $s\sqrt{\hat{\eta}(s)}=s$ is a BF (see the example in Section 2.1.3.). 

\smallskip

{\it {\bf Remark 2.}}
{\it Note that if one uses the sufficient conditions of the subordination approach (see Remark 1), then $s\eta(s)=s$ is CM, but $s^{2}\eta(s)=s^{2}$ is not a BF, thus the sufficient conditions are not fulfilled. However, as we have shown above the solution is non-negative.}

\smallskip

From Eq.~(\ref{second moment}) one easily finds the second moment,
\begin{eqnarray}\label{classical cattaneo eq MSD}
\left\langle x^{2}(t)\right\rangle=t^{2},
\end{eqnarray}
which is an indication of the ballistic motion.

\subsection{Fractional diffusion-wave equation. Power-law memory kernel}

Let us take
\begin{eqnarray}\label{power memory}
\eta(t)=\frac{t^{1-\mu}}{\Gamma(2-\mu)}, \quad \mathrm{i.e.,} \quad \hat{\eta}(s)=s^{\mu-2},
\end{eqnarray}
where $0<\mu<2$. We consider two cases. The case with $1<\mu<2$ corresponds to the fractional wave equation with Caputo fractional derivative (\ref{Caputo_derivative}) of the order $\mu$ \cite{mainardi},
\begin{eqnarray}\label{frac diff mu>1/2}
{_{C}D_{t}^{\mu}}W(x,t)=\frac{\partial^{2}}{\partial x^{2}}W(x,t),
\end{eqnarray}
whereas for the case with $0<\mu<1$ we find
\begin{eqnarray}\label{frac diff mu<1/2}
\frac{1}{\Gamma(2-\mu)}\int_{0}^{t}(t-t')^{1-\mu}\frac{\partial^{2}}{\partial t'^{2}}W(x,t')\,dt'=\frac{\partial^{2}}{\partial x^{2}}W(x,t). 
\end{eqnarray}
We note that the left hand side of Eq.~(\ref{frac diff mu<1/2}) cannot be written as a fractional derivative. The solution of Eqs.~(\ref{frac diff mu>1/2}) and (\ref{frac diff mu<1/2}) in the Fourier-Laplace space is given by
\begin{eqnarray}\label{fractional wave eq LF}
\tilde{\hat{W}}(\kappa,s)=\frac{s^{\mu-1}}{s^{\mu}+\kappa^{2}},
\end{eqnarray}
The solution is non-negative since $\sqrt{\hat{\eta}(s)}=s^{\mu/2-1}$ is CM (see the example in Section 2.1.1.) and $s\sqrt{\hat{\eta}(s)}=s^{\mu/2}$ is a BF for $0<\mu<2$ (see the example in Section 2.1.3.).

\smallskip

{\it {\bf Remark 3.}}
{\it Similar to the previous case of the standard wave equation, here the sufficient conditions of the subordination approach require $s\hat{\eta}(s)=s^{\mu-1}$ to be CM and $s^{2}\hat{\eta}(s)=s^{\mu}$ to be BF. However, both conditions are not satisfied for $1<\mu<2$. }

\smallskip

By the inverse Fourier-Laplace transformation of Eq.~(\ref{fractional wave eq LF}) one finds the solution in terms of the Fox $H$-function,
\begin{eqnarray}\label{fractional wave eq LF one power}
W(x,t)=\mathcal{F}^{-1}\left[E_{\mu}\left(-t^{\mu}\kappa^{2}\right)\right]
=\frac{1}{2|x|}H_{1,1}^{1,0}\left[\left.\frac{|x|}{t^{\mu/2}}\right|\begin{array}{l}\displaystyle\left(1,\mu/2\right)\\(1,1)\end{array}\right].
\end{eqnarray}
Here $E_{\alpha}(z)$ is the one parameter Mittag-Leffler (M-L) function \cite{mainardi_book}
\begin{eqnarray}
E_{\alpha}(z)=\sum_{n=0}^{\infty}\frac{z^n}{\Gamma(\alpha n+1)}, \quad \Re(\alpha)>0,
\end{eqnarray}
and $H_{p,q}^{m,n}(z)$ is the Fox $H$-function which is given in terms of the Mellin transform as \cite{saxena book}
\begin{eqnarray}\label{H_integral}
H_{p,q}^{m,n}(z)&=&H_{p,q}^{m,n}\left[z\left|\begin{array}{c l}
    (a_1,A_1),...,(a_p,A_p)\\
    (b_1,B_1),...,(b_q,B_q)
  \end{array}\right.\right]=H_{p,q}^{m,n}\left[z\left|\begin{array}{c l}
    (a_p,A_p)\\
    (b_q,B_q)
  \end{array}\right.\right]\nonumber\\&=&\frac{1}{2\pi\imath}\int_{\Omega}\theta(s)z^{-s}\textrm{d}s,
\end{eqnarray}
with
$$\theta(s)=\frac{\prod_{j=1}^{m}\Gamma(b_j+B_js)\prod_{j=1}^{n}\Gamma(1-a_j-A_js)}{\prod_{j=m+1}^{q}\Gamma(1-b_j-B_js)\prod_{j=n+1}^{p}\Gamma(a_j+A_js)},$$ $0\leq n\leq p$, $1\leq m\leq q$, $a_i,b_j \in \mathrm{C}$, $A_i,B_j\in\mathrm{R}^{+}$, $i=1,...,p$, $j=1,...,q$. The contour of integration $\Omega$ starts at $c-\imath\infty$ and finishes at $c+\imath\infty$ separating the poles of the function $\Gamma(b_j+B_js)$, $j=1,...,m$ with those of the function $\Gamma(1-a_i-A_is)$, $i=1,...,n$.  

\medskip

{\it {\bf Remark 6.}}
{\it For $\mu=1$ we recover the Gaussian PDF
\begin{eqnarray}\label{gaussian pdf}
W(x,t)=\frac{1}{2|x|}H_{1,1}^{1,0}\left[\left.\frac{|x|}{t^{1/2}}\right|\begin{array}{l}\displaystyle\left(1,1/2\right)\\(1,1)\end{array}\right]=\frac{1}{\sqrt{4\pi t}}e^{-\frac{x^{2}}{4t}}.
\end{eqnarray}
}

\smallskip

The second moment for the fractional diffusion-wave equation (\ref{frac diff mu>1/2}) and (19) is given by 
\begin{eqnarray}\label{MSD 1}
\left\langle x^{2}(t)\right\rangle=2\,\frac{t^{\mu}}{\Gamma(1+\mu)}. 
\end{eqnarray}
Thus, the generalized diffusion-wave equation (\ref{distributed order wave eq memory}) with the memory kernel (\ref{power memory}) is able to describe both superdiffusive and subdiffusive processes since $0<\mu<2$. The case $\mu=1$ reduces to the classical diffusion equation for the Brownian motion, i.e., $\left\langle x^{2}(t)\right\rangle=2\,t$, whereas the case with $\mu=2$ yields ballistic diffusion, $\left\langle x^{2}(t)\right\rangle=t^{2}$.

\subsection{Bi-fractional diffusion-wave equation. Two power-law memory kernels}

Here we consider bi-fractional diffusion-wave equation with the memory kernel of the form
\begin{eqnarray}\label{power memory bi}
\eta(t)=\alpha_{1}\frac{t^{1-\mu_{1}}}{\Gamma(2-\mu_{1})}+\alpha_{2}\frac{t^{1-\mu_{2}}}{\Gamma(2-\mu_{2})}, \quad \alpha_{1}+\alpha_{2}=1,
\end{eqnarray}
and $\hat{\eta}(s)=\alpha_{1}s^{\mu_{1}-2}+\alpha_{2}s^{\mu_{2}-2}$. Again we consider two cases. The case with $1<\mu_1<\mu_2<2$ yields
\begin{eqnarray}\label{two power law}
\alpha_{1}{_C}D_{0+}^{\mu_{1}}W(x,t)+\alpha_{2}{_C}D_{0+}^{\mu_{2}}W(x,t)=\frac{\partial^{2}}{\partial x^{2}}W(x,t),
\end{eqnarray}
where ${_C}D_{0+}^{\mu_{j}}$ is the Caputo fractional derivative (\ref{Caputo_derivative}) of the order $1<\mu_{j}<2$, whereas the case $0<\mu_{1}<\mu_{2}<1$ yields an equation
\begin{eqnarray}\label{distributed order wave eq memory p 2 delta}
&&\frac{\alpha_{1}}{\Gamma(2-\mu_{1})}\int_{0}^{t}(t-t')^{1-\mu_{1}}\frac{\partial^{2}}{\partial t'^{2}}W(x,t')\,dt'\nonumber\\&&+\frac{\alpha_{2}}{\Gamma(2-\mu_{2})}\int_{0}^{t}(t-t')^{1-\mu_{2}}\frac{\partial^{2}}{\partial t'^{2}}W(x,t')\,dt'=\frac{\partial^{2}}{\partial x^{2}}W(x,t).
\end{eqnarray}
The solutions of both equations are non-negative. Let us show this. Since $$\hat{\eta}(s)=\alpha_{1}s^{\mu_1-2}+\alpha_{2}s^{\mu_2-2}$$ is a SF for $1<\mu_{1}<\mu_{2}<2$ as a linear combination of two SFs (see property ({\bf c}) in Section 2.1.2.), then $\sqrt{\hat{\eta}(s)}$ is a SF as a composition of CBF and SF (see property ({\bf g}) in Section 2.1.4.), which means that the function $\sqrt{\hat{\eta}(s)}$ is CM (SFs are subclass of CM functions, see Section 2.1.2.). From here we may conclude that $s\sqrt{\hat{\eta}(s)}$ is a CBF (see property ({\bf h}) in Section 2.1.4.), and hence the solution of Eq.~(\ref{two power law}) is non-negative. On the other hand, for $0<\mu_{1}<\mu_{2}<1$, Eq.~(\ref{distributed order wave eq memory p 2 delta}), we use the subordination approach (see Remark 1), that is $s\hat{\eta}(s)=\alpha_{1}s^{\mu_1-1}+\alpha_{2}s^{\mu_2-1}$ is CM (see property ({\bf a}) and the example in Section 2.1.1.), and $s^{2}\hat{\eta}(s)=\alpha_{1}s^{\mu_1}+\alpha_{2}s^{\mu_2}$ is a BF (see property ({\bf d}) and the example in Section 2.1.3.).

\begin{figure}
\centering
\includegraphics[width=8.5cm]{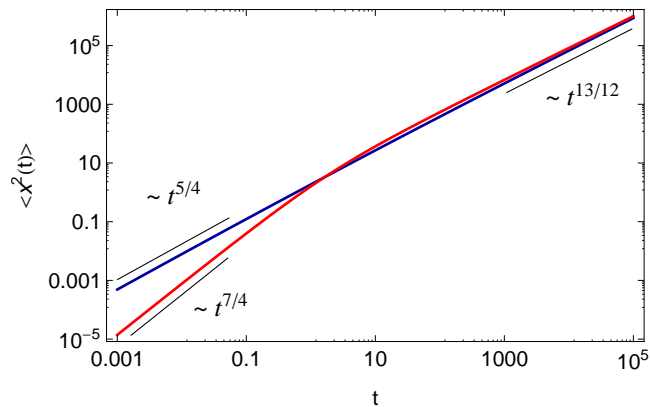} \caption {Graphical representation of the second moment (\ref{msdPower2}) for $a_1=a_2=1/2$, $\mu_1=13/12$ and $\mu_2=5/4$ (blue line), $\mu_2=7/4$ (red line). The crossover from power-law behavior with exponent $\mu_2$ to power-law behavior with an exponent $\mu_1$ is obvious.}\label{figBiFrac}
\end{figure}

The solution of Eqs.~(\ref{two power law}) and (\ref{distributed order wave eq memory p 2 delta}) can be found in terms of infinite series of the Fox $H$-function by using the series expansion approach \cite{podlubny}, as it is done for the bi-fractional diffusion equation in \cite{pre2015}. Therefore, the exact solution becomes
\begin{eqnarray}\label{distributed2 PDF two delta}
P(x,t)&&=\frac{1}{\sqrt{\frac{4\pi}{B_2}t^{\mu_2}}}
\sum_{n=0}^{\infty}\frac{(-1)^n}{n!}\left(\frac{\alpha_1}{\alpha_2}\right)^n t^{\left(\mu_2-\mu_1\right)n}\nonumber\\&&\times\left\{H_{1,2}^{2,0}\left[\frac{x^2}{\frac{4}{\alpha_2}t^{\mu_2}}\left|
\begin{array}{l}\left([\mu_2-\mu_1]n+1-\mu_2/2,\mu_2\right)\\
(0,1),(n+1/2,1)\end{array}\right.\right]\right.\nonumber\\
&&\left.+\frac{\alpha_1}{\alpha_2}t^{\mu_2-\mu_1}H_{1,2}^{2,0}\left[\frac{x^2}{\frac{4}{\alpha_2}t^{\mu_2}}\left|\begin{array}{l}\left([\mu_2-\mu_1](n+1)+1-\mu_2/2,\mu_2\right)\\(0,1),(n+1/2,1)\end{array}\right.\right]\right\},
\end{eqnarray}
while the second moment for both cases, Eqs.~(\ref{two power law}) and (\ref{distributed order wave eq memory p 2 delta}), becomes
\begin{eqnarray}\label{msdPower2}
\left\langle x^{2}(t)\right\rangle=2\,\mathcal{L}^{-1}\left[\frac{s^{-1}}{\alpha_{1}s^{\mu_{1}}+\alpha_{2}s^{\mu_{2}}}\right](t)=\frac{2}{\alpha_{2}}t^{\mu_{2}}E_{\mu_2-\mu_1,\mu_2+1}\left(-\frac{\alpha_{1}}{\alpha_{2}}t^{\mu_{2}-\mu_{1}}\right),\nonumber\\ 
\end{eqnarray}
where $E_{\alpha,\beta}(z)$ is the two parameter M-L function \cite{mainardi_book}
\begin{eqnarray}\label{ml2}
E_{\alpha,\beta}(z)=\sum_{n=0}^{\infty}\frac{z^n}{\Gamma(\alpha n+\beta)}, \quad \Re(\alpha)>0.
\end{eqnarray}
From Eq.~(\ref{ml2}) and by using the formula
\begin{eqnarray}
E_{\alpha,\beta}(-z)\simeq-\sum_{n=1}^{N}\frac{(-z)^{-n}}{\Gamma(\beta-\alpha n)}, \quad z\gg1,
\end{eqnarray}
we obtain the following asymptotics:
\begin{eqnarray}
\label{msd bi-frac asympt}
\langle x^{2}(t)\rangle\simeq\left\lbrace \begin{array}{c l} 
\frac{2}{\alpha_2}\frac{t^{\mu_2}}{\Gamma(1+\mu_2)}, \quad & t\ll1 ,\\
\frac{2}{\alpha_1}\frac{t^{\mu_1}}{\Gamma(1+\mu_1)}, \quad & t\gg1 ,  
\end{array}\right.
\end{eqnarray}
which means {\it decelerating superdiffusion} for $1<\mu_{1}<\mu_{2}<2$, including crossover from superdiffusion to normal diffusion in the case $1=\mu_1<\mu_2<2$, and {\it decelerating subdiffusion} for $0<\mu_{1}<\mu_{2}<1$, including crossover from normal diffusion to subdiffusion for the case $0<\mu_1<\mu_2=1$. Graphical representation of the second moment (\ref{msdPower2}) is given in Figure \ref{figBiFrac}. Such crossover, for example, from superdiffusion to normal diffusion has been observed in Hamiltonian systems with long-range interactions \cite{latora}, and different biological systems \cite{caspi,wun}.

\medskip

{\it {\bf Remark 7.}}
{\it We note that for the case $0<\mu_1<1$ and $1<\mu_2<2$ we are not able to prove the non-negativity of the solution. Thus, whether such diffusion-wave equation for the PDF may exist or not is still an open question.}

\smallskip

\subsection{Fractional diffusion-wave equation with $N$ fractional exponents}\label{app N}

A natural generalization of the previous case is a memory kernel of the form 
\begin{eqnarray}\label{memory N}
\eta(t)=\sum_{j=1}^{N}\alpha_{j}\frac{t^{1-\mu_{j}}}{\Gamma(2-\mu_{j})},
\end{eqnarray}
where $0<\mu_1<\mu_2<\dots<\mu_N<2$. From (\ref{memory N}) we find $\hat{\eta}(s)=\sum_{j=1}^{N}\alpha_{j}s^{\mu_{j}-2}$. Thus, from Eq.~(\ref{distributed order wave eq memory}) with kernel (\ref{memory N}) and $1<\mu_{j}<2$ we get
\begin{eqnarray}\label{distributed order wave eq memory p N delta}
\sum_{j=1}^{N}\alpha_{j}{_C}D_{0+}^{\mu_{j}}W(x,t)=\frac{\partial^{2}}{\partial x^{2}}W(x,t),
\end{eqnarray}
where ${_C}D_{0+}^{\mu_{j}}$ is the Caputo fractional derivative (\ref{Caputo_derivative}) of order $1<\mu_{j}<2$. For $0<\lambda<1$, $0<\mu_{j}<1$, $j=1,2,\dots,N$ we get
\begin{eqnarray}\label{distributed order wave eq memory p N delta2}
\sum_{j=1}^{N}\alpha_{j}\int_{0}^{t}\frac{(t-t')^{1-\mu_{j}}}{\Gamma(2-\mu_{j})}\frac{\partial^{2}}{\partial t'^{2}}W(x,t')\,dt'=\frac{\partial^{2}}{\partial x^{2}}W(x,t).
\end{eqnarray}

The solutions of Eq.~(\ref{distributed order wave eq memory p N delta}) with $1<\mu_1<\mu_2<\dots<\mu_N<2$ is non-negative since $$\hat{\eta}(s)=\sum_{j=1}^{N}\alpha_{j}s^{\mu_j-2}$$ is a SF (see property ({\bf c}) in Section 2.1.2.), that is $\sqrt{\eta(s)}$ a SF too (see property ({\bf g}) in Section 2.1.4.), which means that $\sqrt{\hat{\eta}(s)}$ is CM. Thus, $s\sqrt{\hat{\eta}(s)}$ is a CBF (see property ({\bf h}) in Section 2.1.4.), with which we complete the proof of non-negativity of the solution. For the case with $0<\mu_1<\mu_2<\dots<\mu_N<1$ the non-negativity of the solution of Eq.~(\ref{distributed order wave eq memory p N delta2}) can be shown by using the subordination approach (see Remark 1), that is $s\hat{\eta}(s)=\sum_{j=1}^{N}\alpha_{j}s^{\mu_j-1}$ a CM, and $s^{2}\hat{\eta}(s)=\sum_{j=1}^{N}\alpha_{j}s^{\mu_j}$ is a BF.

For the MSD we find
\begin{eqnarray}\label{msdPowerN}
&&\left\langle x^{2}(t)\right\rangle=2\,\mathcal{L}^{-1}\left[\frac{s^{-1}}{\sum_{j=1}^{N}\alpha_{j}s^{\mu_{j}}}\right](t)\nonumber\\&&=\frac{2}{\alpha_{N}}t^{\mu_{N}}E_{(\mu_N-\mu_1,\dots,\mu_N-\mu_{N-1}),\mu_N+1}\left(-\frac{\alpha_{1}}{\alpha_{N}}t^{\mu_{N}-\mu_{1}},\dots,-\frac{\alpha_{N-1}}{\alpha_{N}}t^{\mu_{N}-\mu_{N-1}}\right),\nonumber\\ 
\end{eqnarray}
where $E_{(a_1,a_2,\dots,a_N),b}(z)$ is the multinomial M-L function \cite{luchko_gorenflo,luchko h t}, defined by 
\begin{eqnarray}
\fl E_{\left(a_{1},a_{2},\ldots,a_{N}\right),b}\left(z_{1},z_{2},\ldots,z_{N}\right)=
\sum_{j=1}^{\infty}\sum_{k_{1}\geq0,k_{2}\geq0,\ldots,k_{N}\geq0}^{k_{1}+k_{2}+
\ldots+k_{N}=j}&&\left(\begin{array}{c}j\\k_{1} \quad k_{2} \quad \ldots \quad k_{N}
\end{array}\right)\nonumber\\&&\times\frac{\prod_{i=1}^{N}\left(z_{i}\right)^{k_{i}}}{\Gamma\left(b+
\sum_{i=1}^{N}a_{i}k_{i}\right)},
\end{eqnarray}
where $$\left(\begin{array}{c}j\\k_{1} \quad k_{2} \quad ... \quad k_{N}\end{array}\right)=
\frac{j!}{k_{1}!k_{2}!...k_{N}!}$$ are the so-called multinomial coefficients. For the asymptotic behavior of the second moment we get
\begin{eqnarray}
\label{msd N-frac asympt}
\langle x^{2}(t)\rangle\simeq\left\lbrace \begin{array}{c l} 
\frac{2}{\alpha_N}\frac{t^{\mu_N}}{\Gamma(1+\mu_N)}, \quad & t\ll1\\
\frac{2}{\alpha_1}\frac{t^{\mu_1}}{\Gamma(1+\mu_1)}, \quad & t\gg1
\end{array}\right.
\end{eqnarray}
i.e., decelerating superdiffusion for $1<\mu_1<\mu_2<\dots<\mu_N<2$, and decelerating subdiffusion for $0<\mu_1<\mu_2<\dots<\mu_N<1$. We note that the multinomial M-L function plays an important role in the analysis of the generalized Langevin equation \cite{sandev pla2014} and distributed order diffusion equations \cite{fcaa2015,pre2015}. 

\begin{figure}
\centering
\includegraphics[width=8.5cm]{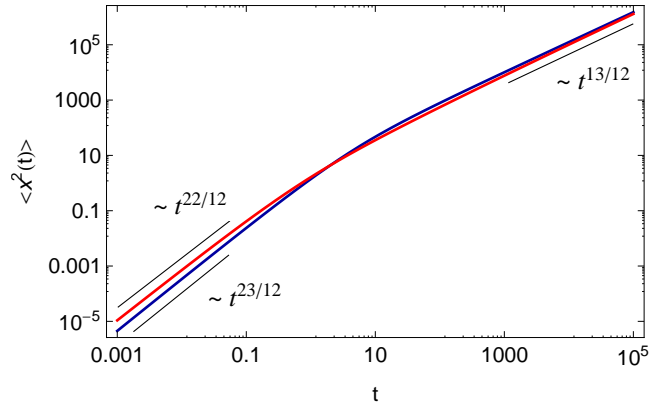} \caption {Graphical representation of the second moment (\ref{msdPowerN}) for $a_1=a_2=a_3=1/3$, $\mu_1=13/12$, $\mu_2=21/12$, $\mu_3=23/12$ (blue line), $\mu_1=13/12$, $\mu_2=15/12$, $\mu_3=22/12$ (red line).}\label{msdPower3}
\end{figure}

Graphical representation of the second moment in case of three power-law memory kernels ($N=3$) is given in Figure \ref{msdPower3}. It is seen that the second moment in the short time limit behaves as power-law with exponent $\mu_3$, which turns to power-law behavior with an exponent $\mu_1$. In the intermediate time the behavior is represented by the multinomial M-L function. By introducing more than two power-law memory kernels it is possible to achieve better fit in the intermediate time domain between short and long time asymptotics.

\subsection{Uniformly distributed order memory kernel}

Next we consider uniformly distributed memory kernel $$\eta(t)=\int_{1}^{2}\frac{t^{1-\lambda}}{\Gamma(2-\lambda)}\,d\lambda,$$ that corresponds to $p(\lambda)=1$ in Eq.~(\ref{distributed order wave eq}). Equation~(\ref{distributed order wave eq memory}) then reads
\begin{eqnarray}\label{distributed order wave eq memory p=1}
\int_{0}^{t}\int_{1}^{2}\frac{(t-t')^{1-\lambda}}{\Gamma(2-\lambda)}\frac{\partial^{2}}{\partial t'^{2}}W(x,t')\,d\lambda {d}t'=\frac{\partial^{2}}{\partial x^{2}}W(x,t),
\end{eqnarray}
i.e.,
\begin{eqnarray}\label{distributed order wave eq memory p=1 2}
\int_{1}^{2}{_C}D_{0+}^{\lambda}W(x,t)\,d\lambda=\frac{\partial^{2}}{\partial x^{2}}W(x,t),
\end{eqnarray}
where ${_C}D_{0+}^{\lambda}$ is the Caputo fractional derivative of the order $1<\lambda<2$. 

Let us show the non-negativity of the solution. The function $$\hat{\eta}(s)=\int_1^2p(\lambda)s^{\lambda-2}\,d\lambda$$ is a SF since the function $\sum_{j}p_{j}s^{\lambda_{j}-2}$ with $p_{j}\geq0$ and $1<\lambda_{j}\leq2$ is a SF (see property ({\bf c}) in Section 2.1.2.), and a pointwise limit of this linear combination $\eta(s)=\int_0^1p(\lambda)s^{\lambda-2}\,d\lambda$ is a SF too. Therefore, $\sqrt{\hat{\eta}(s)}$ is a SF as well, since a composition of CBF and SF is a SF (see property ({\bf g}) in Section 2.1.4.), and thus it is CM. Therefore, $s\sqrt{\hat{\eta}(s)}$ is CBF (if $c(x)$ is a CBF, then $c(x)/x$ is a SF (see property ({\bf h}) in Section 2.1.4.), which means that the solution of the equation (\ref{distributed order wave eq memory p=1 2}) is non-negative.

From the solution in Laplace space,
\begin{eqnarray}\label{asympt pdf}
\hat{W}(x,s)=\frac{1}{2s}\left(s\sqrt{\frac{s-1}{s\log{s}}}\right)\exp\left(-s\sqrt{\frac{s-1}{s\log{s}}}|x|\right),
\end{eqnarray}
by applying the Tauberian theorem for slowly varying functions ~\cite{feller} we can obtain the asymptotic behavior of $W(x,t)$ in the short and long time limit. This theorem states that if some function $r(t)$, $t\geq 0$, has the Laplace transform $\hat{r}(s)$ whose asymptotics behaves as
\begin{eqnarray}\label{tauber7}
\hat{r}(s)\simeq s^{-\rho}L\left(\frac{1}{s}\right), \quad
s\rightarrow0, \quad \rho\geq0,
\end{eqnarray}
then
\begin{eqnarray}\label{tauber8}
r(t)=\mathcal{L}^{-1}\left[\hat{r}(s)\right]\simeq
\frac{1}{\Gamma(\rho)}t^{\rho-1}L(t), \quad t\rightarrow\infty.
\end{eqnarray}
Here $L(t)$ is a slowly varying function at infinity, i.e., $$\lim_{t\rightarrow\infty}\frac{L(at)}{L(t)}=1,$$ for any $a>0$. The theorem is also valid if $s$ and $t$ are interchanged, that is $s\rightarrow\infty$ and $t\rightarrow0$ \cite{feller}. The value of $\rho$ is equal to $0$ and $1/2$ for short and long time limit respectively. Therefore, from Eq.~(\ref{asympt pdf}) we find
\begin{eqnarray}
W(x,t)\simeq\left\lbrace \begin{array}{c l} 
\frac{1}{\sqrt{4t^{2}\log{\frac{1}{t}}}}\exp\left(-\frac{|x|}{\sqrt{t^{2}\log{\frac{1}{t}}}}\right), \quad & t\ll1,\\
\frac{1}{\sqrt{4t\log{t}}}\exp\left(-\frac{|x|}{\sqrt{t\log{t}}}\right), \quad & t\gg1.
\end{array}\right.
\end{eqnarray}
The MSD reads
\begin{eqnarray}\label{msdD}
\left\langle x^{2}(t)\right\rangle=2\,\mathcal{L}^{-1}\left[\frac{\log{s}}{s^{2}(s-1)}\right](t). 
\end{eqnarray}
From here, by applying the Tauberian theorem for slowly varying functions, we obtain
\begin{eqnarray}
\label{msd distr asympt}
\langle x^{2}(t)\rangle\simeq\left\lbrace \begin{array}{c l} 
2\,\mathcal{L}^{-1}\left[s^{-3}\log{s}\right](t)\simeq t^{2}\log\frac{1}{t}, \quad & t\ll1,\\
2\,\mathcal{L}^{-1}\left[s^{-2}\log\frac{1}{s}\right](t)=2\, t\left(-1+\gamma+\log{t}\right)\simeq2\,t\log{t}, \quad & t\gg1.
\end{array}\right.\nonumber\\
\end{eqnarray}
Therefore, the behavior of the MSD is {\it weakly superballistic} at short times and {\it weakly superdiffusive} at long times. Interestingly, the MSD (\ref{msd distr asympt}) for the diffusion--wave equation (\ref{distributed order wave eq memory p=1 2}) in comparison to the MSD for the diffusion equation with uniformly distributed order memory kernel \cite{chechkin,chechkin2}, has an additional multiplicative term $t$ for both short and long times.

\subsection{Truncated power-law memory kernel}

Next we present results in case of a truncated power-law memory kernel of the form
\begin{eqnarray}\label{tr kernel}
\eta(t)=e^{-bt}\frac{t^{1-\mu}}{\Gamma(2-\mu)},
\end{eqnarray}
where $b>0$, and $1\le\mu<2$. Its Laplace transform is given by $\hat{\eta}(s)=(s+b)^{\mu-2}$, where we use the shift rule of the Laplace transform 
\begin{eqnarray}\label{shift rule}
\mathcal{L}\left[f(t)e^{-at}\right](s)=\hat{F}
(s+a), \quad \mathrm{with} \quad \mathcal{L}\left[f(t)\right](s)=\hat{F}(s).
\end{eqnarray}
The solution of the diffusion-wave equation with the truncated memory kernel (\ref{tr kernel}) is non-negative for $1\le\mu<2$ since $\hat{\eta}(s)=(s+b)^{\mu-2}$ is a SF as a composition of SF ($s^{\mu-2}$, $1\leq\mu<2$) and CBF ($s+b$) (see property ({\bf f}) in Section 2.1.4.). Therefore $\sqrt{\eta(s)}$ is a SF too, with which we complete the proof. 

From the solution in Laplace space we find the exact solution in the following form
\begin{eqnarray}
W(x,t)&=&\mathcal{L}^{-1}\left[\frac{1}{2s}(s+b)^{\mu/2}\exp\left(-(s+b)^{\mu/2}|x|\right)\right](x,t)\nonumber\\&=&{_{RL}}I_{t}^{1}\left(\frac{e^{-bt}}{2|x|t}H_{1,1}^{1,0}\left[\frac{|x|}{t^{\mu/2}}\left|\begin{array}{c l}
    (0,\mu/2)\\
    (0,1)
  \end{array}\right.\right]\right),
\end{eqnarray}
where we use the relation between the exponential and the Fox $H$-function \cite{saxena book}
\begin{eqnarray}\label{exp H}
e^{-z}=H_{0,1}^{1,0}\left[z\left|\begin{array}{l}
    -\\
    (0,1)
  \end{array}\right.\right],
\end{eqnarray}
the shift rule (\ref{shift rule}), and the Laplace transform formula \cite{saxena book}
\begin{eqnarray}\label{Laplace H}
\mathcal{L}^{-1}\left[s^{-\rho}H_{p,q}^{m,n}\left[zs^{\sigma}\left|\begin{array}{c
l}
    (a_p,A_p)\\
    (b_q,B_q)
  \end{array}\right.\right]\right](t)=t^{\rho-1}H_{p+1,q}^{m,n}\left[zt^{-\sigma}\left|\begin{array}{c
  l}
      (a_p,A_p),(\rho,\sigma)\\
      (b_q,B_q)
    \end{array}\right.\right],\nonumber\\
\end{eqnarray}
where $\sigma>0$, $\Re(s)>0$, $\Re\left(\rho+\sigma \max_{1\leq j\leq n}\left(\frac{1-a_j}{A_j}\right)\right)>0$, $|\arg(z)|<\pi\theta_1/2$, $\theta_1>0$, $\theta_1=\theta-a$.

From Eq. (\ref{second moment}) for the second moment we obtain
\begin{eqnarray}
\left\langle x^{2}(t)\right\rangle=2\,\mathcal{L}^{-1}\left[s^{-3}(s+b)^{2-\mu}\right](t)=2\,{_{RL}}I_{t}^{3}\left(e^{-bt}\frac{t^{-3+\mu}}{\Gamma(-2+\mu)}\right), 
\end{eqnarray}
where 
\begin{eqnarray}\label{RLintegral}
{_{RL}I_t^{\alpha}}f(t)=\frac{1}{\Gamma(\alpha)}\int_0^t(t-t')^{\alpha-1}f(t')\,dt'
\end{eqnarray}
is the Riemann-Liouville integral, whose Laplace transform is given by $\mathcal{L}\left[{_{RL}I_t^{\alpha}}f(t)\right](s)=s^{-\alpha}\mathcal{L}\left[f(t)\right](s)$ \cite{podlubny}. For the asymptotic behavior of the second moment we use the Tauberian theorem \cite{feller}, which states that if the asymptotic behavior of $r(t)$ for $t\rightarrow\infty$ is given by
\begin{eqnarray}\label{tauber1}
r(t)\simeq t^{-\alpha}, \quad t\rightarrow\infty,
\end{eqnarray}
then, the corresponding Laplace pair $\hat{r}(s)=\mathcal{L}[r(t)](s)$ has the following behavior for $s\rightarrow0$
\begin{eqnarray}\label{tauber2}
\hat{r}(s)\simeq \Gamma(1-\alpha)s^{\alpha-1}, \quad s\rightarrow0,
\end{eqnarray}
and vice-versa, ensuring that $r(t)$ is non-negative and monotone function at infinity \cite{feller}. Thus, we find 
\begin{eqnarray}
\label{msd trunc asympt}
\langle x^{2}(t)\rangle\simeq\left\lbrace \begin{array}{c l} 
2\,\frac{t^{\mu}}{\Gamma(1+\mu)}, \quad & t\ll1,\\
b^{2-\mu}t^{2}, \quad & t\gg1,
\end{array}\right.
\end{eqnarray}
which means that the process switches from superdiffusive behavior to ballistic motion in the case with $1<\mu<2$, and from normal diffusion to ballistic motion in the case with $\mu=1$. The fact that we encounter accelerating superdiffusion caused by truncation of the power law kernel in the {\it diffusion-wave equation} is not surprising. Indeed, it is analogous to the transition from subdiffusion to normal diffusion described by the fractional {\it diffusion equation} with truncated power law kernel in the Caputo form \cite{fcaa2015,csf2017}. As in diffusion equation the truncation naturally leads to a normal diffusion at long times, in diffusion--wave equation the truncation results in the long--time ballistic behavior.

\subsection{Truncated Mittag-Leffler memory kernel}

Here we consider tempered M-L memory kernel of form
\begin{eqnarray}\label{gamma tempered ml}
\eta(t)=e^{-bt}t^{\beta-1}E_{\alpha,\beta}^{\delta}\left(-\nu t^{\alpha}\right),
\end{eqnarray}
where $\alpha\delta<\beta<1$, and $E_{\alpha,\beta}^{\delta}(z)$ is the three parameter M-L function \cite{prabhakar} (for details on three parameter M-L function we refer to \cite{garrappa4,Tomovski2015})
\begin{eqnarray}\label{three ml}
E_{\alpha,\beta}^{\delta}(z)=\sum_{k=0}^{\infty}\frac{(\delta)_k}{\Gamma(\alpha k+\beta)}\frac{z^k}{k!},
\end{eqnarray}
$\beta, \delta, z \in \mathrm{C}$, $\Re(\alpha)>0$, and $(\delta)_{k}$ is the Pochhammer symbol $(\delta)_{0}=1$, $(\delta)_{k}=\frac{\Gamma(\delta+k)}{\Gamma(\delta)}$. The asymptotic behavior of the three parameter M-L function for large arguments follows from the formula \cite{garrappa4,pre2015} 
\begin{eqnarray}\label{ml three asympt}
E_{\alpha,\beta}^{\delta}(-t^{\alpha})=\frac{t^{-\alpha\delta}}{\Gamma(\delta)}\sum_{n=0}^{\infty}\frac{\Gamma(\delta+n)}{\Gamma(\beta-\alpha(\delta+n)}\frac{(-t^{\alpha})^{-n}}{n!}, \quad t\gg1,
\end{eqnarray}
for $0<\alpha<2$. The Laplace transform of the memory kernel becomes $$\hat{\eta}(s)=\frac{(s+b)^{\alpha\delta-\beta}}{\left[(s+b)^{\alpha}+\nu\right]^{\delta}},$$ where we use the Laplace transform formula \cite{prabhakar}
\begin{eqnarray}\label{Laplace ML3_1}
\mathcal{L}\left[t^{\beta-1}E_{\alpha,\beta}^{\delta}\left(-\lambda t^{\alpha}\right)\right](s)=\frac{s^{\alpha\delta-\beta}}{(s^\alpha+\lambda)^\delta}, \quad |\lambda/s^{\alpha}|<1.
\end{eqnarray} 
Therefore, the generalized diffusion-wave equation has the form
\begin{eqnarray}\label{wave tempered ml}
\int_{0}^{t}(t-t')^{\beta-1}E_{\alpha,\beta}^{\delta}\left(-\nu[t-t']^{\alpha}\right)\frac{\partial^{2}}{\partial t'^{2}}W(x,t')\,dt'=\frac{\partial^2}{\partial x^2}W(x,t).
\end{eqnarray}
In order to find the parameters' constraints, for which the solution of Eq.~(\ref{wave tempered ml}) is non-negative, we notice that $\hat{\eta}(s)$ is a SF (and thus CM) if both functions $(s+b)^{-(\beta-\alpha\delta)}$ and $\left[(s+b)^{\alpha}+\nu\right]^{-\delta}$ are SFs. The first one is a SF if $0<\beta-\alpha\delta<1$ and the second one -- if $0<\alpha\delta<1$. Therefore, we use $0<\alpha\delta<\beta<1$.

For the MSD we find
\begin{eqnarray}\label{second moment tempered ml}
\left\langle
x^{2}(t)\right\rangle&=&2\,\mathcal{L}^{-1}\left[s^{-3}\frac{(s+b)^{-\alpha\delta+\beta}}{\left[(s+b)^{\alpha}+\nu\right]^{-\delta}}\right](t)\nonumber\\
&=&2\,{_{RL}}I_{t}^{3}\left(e^{-bt}t^{-\beta-1}E_{\alpha,-\beta}^{-\delta}\left(-\nu t^{\alpha}\right)\right).
\end{eqnarray}
Again, by using the Tauberian theorem \cite{feller}, for the short and long time limit we find 
\begin{eqnarray}
\label{msd distr ml asympt}
\langle x^{2}(t)\rangle\simeq\left\lbrace \begin{array}{c l} 
2\,t^{2-\beta}/\Gamma(3-\beta), \quad & t\ll1,\\
b^{-\alpha\delta+\beta}\left(b^{\alpha}+\nu\right)^{\delta}t^{2}, \quad & t\gg1.
\end{array}\right.
\end{eqnarray}
Therefore, truncation (tempering) of the M-L kernel causes ballistic motion in the long time limit, which follows the superdiffusive initial behavior. Here we note that tempered M-L memory kernel has been introduced in the analysis of the generalized Langevin equation \cite{sandev liemert}, where the crossover from initial subdiffusion to normal diffusion is observed. The case with no truncation ($b=0$) for the second moment yields $\left\langle
x^{2}(t)\right\rangle=2t^{2-\beta}E_{\alpha,3-\beta}^{-\delta}\left(-\nu t^{\alpha}\right)$, which in the long time limit behaves as $\left\langle
x^{2}(t)\right\rangle\simeq t^{2+\alpha\delta-\beta}$.

\begin{figure}
\centering
\includegraphics[width=8.5cm]{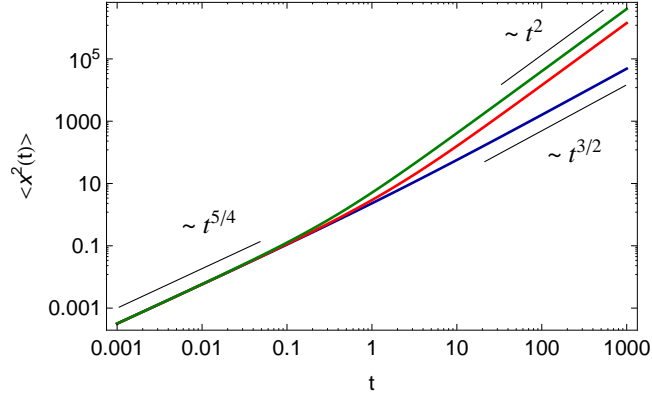} \caption {Graphical representation of the second moment (\ref{second moment tempered ml}) for the case of truncated M-L memory kernel for $\alpha=\delta=1/2$, $\beta=3/4$, $\nu=1$ and $b=0$ (blue line), $b=1$ (red line), $b=5$ (green line).}\label{figTML}
\end{figure}

\begin{figure}
\centering
\includegraphics[width=8.5cm]{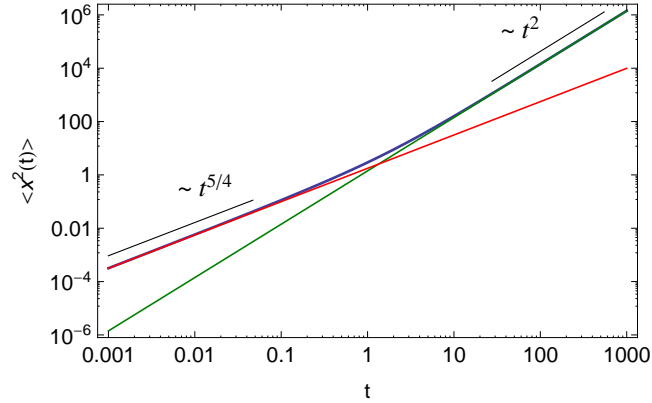} \caption {Graphical representation of (\ref{second moment tempered ml}) for $\alpha=\delta=1/2$, $\beta=3/4$, $\nu=1$, $b=1$ (blue line). The short time limit $\left\langle
x^{2}(t)\right\rangle\simeq2\,t^{2-\beta}/\Gamma(3-\beta)$ (red line) and long time limit $\left\langle
x^{2}(t)\right\rangle\simeq t^{2}b^{-\alpha\delta+\beta}\left(b^{\alpha}+\nu\right)^{\delta}$ (green line) are in excellent agreement with the exact result.}\label{figTML2}
\end{figure}

Graphical representation of the second moment (\ref{second moment tempered ml}) is given in Figure \ref{figTML}. The influence of tempering on particle behavior is clearly observed. Comparison with the asymptotic behaviors in the short and long time limit is given in Figure \ref{figTML2}.

\subsection{Case with regularized Prabhakar derivative}

Let us now consider the regularized Prabhakar derivative defined by \cite{garra}
\begin{eqnarray}\label{prabhakar derivative}
{_C}\mathcal{D}_{\rho,-\nu,t\,}^{\delta,\mu}f(t)=\left(\mathcal{E}_{\rho,m-\mu,-\nu,t\,}^{-\delta}\frac{d^{m}}{dt^{m}}f\right)(t),
\end{eqnarray}
where $\mu, \nu, \delta, \rho \in C$, $\Re(\mu)>0$, $\Re(\rho)>0$, $m=[\mu]+1$, and
\begin{eqnarray}\label{prabhakar integral}
\left(\mathcal{E}_{\rho,m-\mu,-\nu,t\,}^{-\delta}f\right)(t)=\int_{0}^{t}(t-t')^{m-\mu-1}E_{\rho,m-\mu}^{-\delta}\left(-\nu(t-t')^{\rho}\right)f(t')\,dt'
\end{eqnarray}
is the Prabhakar integral \cite{prabhakar}. Therefore, for $1<\mu<2$ ($m=2$) we have
\begin{eqnarray}\label{prabhakar derivative m=2}
\fl{_C}\mathcal{D}_{\rho,-\nu,t\,}^{\delta,\mu}f(t)&=&\left(\mathcal{E}_{\rho,2-\mu,-\nu,t\,}^{-\delta}\frac{d^{2}}{dt^{2}}f\right)(t)=\int_{0}^{t}(t-t')^{1-\mu}E_{\rho,2-\mu}^{-\delta}\left(-\nu(t-t')^{\rho}\right)\frac{d^2}{dt'^2}f(t')\,dt'\nonumber\\&=&\int_{0}^{t}\eta(t-t')\frac{d^2}{dt'^2}f(t')\,dt',
\end{eqnarray}
where
\begin{eqnarray}\label{gamma prabhakar}
\eta(t)=t^{1-\mu}E_{\rho,2-\mu}^{-\delta}\left(-\nu t^{\rho}\right), \quad 1<\mu<2,
\end{eqnarray}
that is the kernel on the left hand side of Eq.~(\ref{distributed order wave eq memory}). This resembles the case considered in subsection III.G, see Eq.~(\ref{gamma tempered ml}), if $b=0$, however, in that previous case the MSD asymptotics at long times is determined by $b$, see Eq.~(\ref{msd distr ml asympt}). This is why the kernel (\ref{gamma prabhakar}) is considered separately. We also use that $0<\rho<1$ and $0<\delta<1$. The generalized diffusion-wave equation then becomes
\begin{eqnarray}\label{wave Pr}
{_C}\mathcal{D}_{\rho,-\nu,t\,}^{\delta,\mu}W(x,t)=\frac{\partial^2}{\partial x^2}W(x,t).
\end{eqnarray}

The Prabhakar derivative has been introduced and applied to fractional Poisson processes in \cite{garra}, and further used in viscoelastic theory \cite{garrappa,garrappa2,giusti}, fractional differential filtration dynamics \cite{bulavatsky}, and generalized Langevin equation \cite{mathematics2017}. Moreover, the fractional diffusion equation with regularized Prabhakar derivative has been recently derived from the continuous time random walk theory \cite{sandev_jpa2018}.

The non-negativity of the solution of Eq.~(\ref{wave Pr}) can be proven as follows. The function $$\hat{\eta}(s)=\left(s^{\frac{\mu-2}{\delta}}+\nu s^{\frac{\mu-2}{\delta}-\rho}\right)^{\delta}$$ is a SF if $s^{(\mu-2)/\delta}$ and $s^{(\mu-2)/\delta-\rho}$ are SFs, and $0<\delta<1$ (composition of a CBF and SF is a SF, see property ({\bf g}) in Section 2.1.4.). Therefore, we obtain that $0<2-\mu<\delta$ and $2-\delta<\mu-\rho\delta<2$. For these values of parameters $\sqrt{\hat{\eta}(s)}$ is a SF as well, with which we show the non-negativity of the solution.

Thus, for the second moment we obtain
\begin{eqnarray}\label{second moment Pr}
\left\langle
x^{2}(t)\right\rangle&=&2\,\mathcal{L}^{-1}\left[\frac{s^{-3}}{s^{-\rho\delta+\mu-2}(s^{\rho}+\nu)^{\delta}}\right](t)
=2\,\mathcal{L}^{-1}\left[\frac{s^{-1+\rho\delta-\mu}}{(s^{\rho}+\nu)^{\delta}}\right](t)\nonumber\\&=&2\,t^{\mu}E_{\rho,\mu+1}^{\delta}\left(-\nu t^{\rho}\right).
\end{eqnarray}
From here, by using Eq.~(\ref{ml three asympt}), we find the asymptotic behavior in the short and long time limit,
\begin{eqnarray}
\label{msd prabh asympt}
\langle x^{2}(t)\rangle\simeq\left\lbrace \begin{array}{c l} 
2\,t^{\mu}/\Gamma(1+\mu), \quad & t\ll1,\\
2\,t^{\mu-\rho\delta}/\Gamma(1+\mu-\rho\delta), \quad & t\gg1,
\end{array}\right.
\end{eqnarray}
which implies that the random motion shows decelerating superdiffusive behavior. %switches from superdiffusive behavior to either superdiffusive, normal or subdiffusive behavior.

\subsection{Distributed order regularized Prabhakar derivative}

Further, for the first time in the literature we introduce the distributed order M-L memory kernel of the form
\begin{eqnarray}\label{gamma prabhakar distributed}
\eta(t)=\int_{1}^{2}t^{1-\mu}E_{\rho,2-\mu}^{-\delta}\left(-\nu t^{\rho}\right)\,d\mu,
\end{eqnarray}
where $0<\mu,\rho,\delta<1$, $0<2-\mu<\delta$ and $2-\delta<\mu-\rho\delta<2$. The generalized diffusion-wave equation becomes distributed order wave equation with regularized Prabhakar derivative,
\begin{eqnarray}\label{wave Pr distributed}
\int_{1}^{2}{_C}\mathcal{D}_{\rho,-\nu,t\,}^{\delta,\mu}W(x,t)\,d\mu=\frac{\partial^2}{\partial x^2}W(x,t).
\end{eqnarray}
Note that for $\delta=0$ Eq.~(\ref{wave Pr distributed}) is reduced to the uniformly distributed order wave equation (\ref{distributed order wave eq memory p=1 2}).

From the memory kernel (\ref{gamma prabhakar distributed}) we find that 
\begin{eqnarray}
\hat{\eta}(s)=\frac{s-1}{s\log{s}}\left(1+\frac{\nu}{s^{\rho}}\right)^{\delta}.
\end{eqnarray}

In order to check the non-negativity of the solution of Eq.~(\ref{wave Pr distributed}), we should analyze the function $$\hat{\eta}(s)=\int_1^2\left(s^{\frac{\mu-2}{\delta}}+\nu s^{\frac{\mu-2}{\delta}-\rho}\right)^{\delta}\,d\mu.$$ It is a SF for the same restrictions of parameters as those for the diffusion-wave equation with regularized Prabhakar derivative since a pointwise limit of the linear combinations of SFs, $$\sum_{j}\left(s^{\frac{\mu_j-2}{\delta}}+\nu s^{\frac{\mu_j-2}{\delta}-\rho}\right)^{\delta}$$ is also SF.

For the MSD we obtain
\begin{eqnarray}\label{second moment Pr distributed}
\left\langle
x^{2}(t)\right\rangle=2\,\mathcal{L}^{-1}\left[\frac{\log{s}}{s^2(s-1)}\left(1+\frac{\nu}{s^{\rho}}\right)^{-\delta}\right](t).
\end{eqnarray}
Therefore, from the Tauberian theorem the asymptotics at short and long times read
\begin{eqnarray}
\label{msd distr prabh asympt}
\langle x^{2}(t)\rangle\simeq\left\lbrace \begin{array}{c l} 
2\,\mathcal{L}^{-1}\left[\frac{\log{s}}{s^3}\right](t)\simeq t^{2}\log\frac{1}{t}, \quad & t\ll1,\\
\frac{2}{\nu^\delta}\mathcal{L}^{-1}\left[\frac{\log{\frac{1}{s}}}{s^{2-\rho\delta}}\right](t)\simeq\frac{2}{\nu^\delta}\frac{t^{1-\rho\delta}}{\Gamma(2-\rho\delta)}\log{t}, \quad & t\gg1,
\end{array}\right.
\end{eqnarray}
and thus, the random motion is {\it weakly superballistic} at short times and weakly subdiffusive at long times since $0<1-\rho\delta<1$.

\section{Discussion: The role of the non-zero second initial condition for the generalized diffusion-wave equation (\ref{distributed order wave eq memory})}

Let us now analyze the solution of the generalized diffusion-wave equation (\ref{distributed order wave eq memory}), which we denote as $W^\ast(x,t)$, with non-zero second initial condition, for example, $\frac{\partial}{\partial t}W^\ast(x,0)=\varphi(x)$, and $W^\ast(x,0)=\delta(x)$. The solution in the Fourier-Laplace space reads
\begin{eqnarray}\label{distributed order wave eq Laplace-Fourier space M 2}
\tilde{\hat{W}}^\ast(\kappa,s)=\frac{s\hat{\eta}(s)}{s^{2}\hat{\eta}(s)+\kappa^{2}}+\frac{1}{s}\frac{s\hat{\eta}(s)}{s^{2}\hat{\eta}(s)+\kappa^{2}}\tilde{\varphi}(\kappa)=\hat{W}(\kappa,s)\left[1+s^{-1}\tilde{\varphi}(\kappa)\right],\nonumber\\
\end{eqnarray}
where $\tilde{\varphi}(\kappa)=\mathcal{F}\left[\varphi(x)\right]$. Therefore, by inverse Fourier transformation we find
\begin{eqnarray}\label{PDF L 2}
\hat{W}^\ast(x,s)=W(x,s)+s^{-1}\int_{-\infty}^{\infty}W(x-\xi,s)\varphi(\xi)\,d\xi,
\end{eqnarray}
from which by inverse Laplace transform we find the exact solution
\begin{eqnarray}\label{PDF L 2}
W^\ast(x,t)=W(x,t)+\int_{0}^{t}\left[\int_{-\infty}^{\infty}W(x-\xi,t)\varphi(\xi)\,d\xi\right]dt'.
\end{eqnarray}
From Eq.~(\ref{distributed order wave eq Laplace-Fourier space M 2}) we find that $$\int_{-\infty}^{\infty}W(x,t)\,dx=1+\tilde{\varphi}(\kappa=0)\,t,$$ therefore, in order for $W(x,t)$ to be normalized to $1$, we need to have $\tilde{\varphi}(\kappa=0)=0$. 

Furthermore, for the the second moment we obtain
\begin{eqnarray}\label{second moment nonzero initial cond}
\left\langle
x^{2}(t)\right\rangle=2\,\mathcal{L}^{-1}\left[\frac{1}{s^{3}\hat{\eta}(s)}\right](t)-\tilde{\varphi}''(\kappa=0)\,t.
\end{eqnarray}
By comparison with the MSD given by Eq.~(13) we see that the additional term appears which is linear in time. From the previous analysis of the second moment for zero second initial condition, this term is negligible in the long time limit in case of superdiffusion, and in the short time limit in case of subdiffusion, in the corresponding model with zero second initial condition. It also follows from Eq.~(80) that there is a restriction such that the obtained second moment from the PDF must be positive. We also note that additional restrictions will appear from the requirement that all even moments must be non-negative $\left\langle x^{2n}(t)\right\rangle\geq0$. Thus, the case with the  non-zero second initial condition looks non-trivial, and it is reasonable to address this issue for each particular physical situation separately.

\section{Summary}

We study generalized diffusion-wave equation with a general memory kernel, which brings as a particular case the distributed order time fractional diffusion-wave equation considered earlier. We give the general form of the solution and find the conditions under which it is non-negative for various particular model forms of the memory kernel. For these kernels we also calculated the mean squared displacements and show that the models suggested may describe multi-scaling diffusion processes characterized by different time exponents for different time intervals. The results are summarized in Table \ref{table}. We thus obtain a flexible tool which can be applied for the description of diverse diffusion phenomena in complex systems, which demonstrate a non-monoscaling behavior, that is transitions between different diffusion regimes. Of course, in each particular application of these models the physical background should be discussed, for example, the existence of a relevant continuous time random walk description, or the particular form of the second initial condition as discussed in Section 4. We note that a very different model based on the Langevin equation with generalized memory kernel has been studied very recently, which also leads to  different crossovers between the diffusion regimes \cite{njp2018}.

\begin{table}[h]
\centering{\caption{Behavior of the MSD in the short and long time limit obtained from the generalized diffusion-wave equation (\ref{distributed order wave eq memory}), $\int_{0}^{t}\eta(t-t')\frac{\partial^2}{\partial t'^2}W(x,t')\,\mathrm{d}t'=\frac{\partial^{2}}{\partial x^{2}}W(x,t)$, for different forms of the memory kernel $\eta(t)$.}\label{table}}
\begin{tabular}{|l|l|l|l|l|}
\hline
     $\quad \eta(t) \quad$       & $\quad \left\langle x^{2}(t)\right\rangle, \quad t\ll1 \quad $       & $\quad \left\langle x^{2}(t)\right\rangle, \quad t\gg1 \quad $ \\ \hline
  $\quad \delta(t) \quad$       & $\quad \sim t^2$       & $\quad \sim t^2$ \\ \hline
  $\quad \frac{t^{1-\mu}}{\Gamma(2-\mu)}, \quad 0<\mu<2 \quad$       & $\quad \sim t^\mu$       & $\quad \sim t^\mu$ \\ \hline
  $\begin{array}{c l} & \alpha_1\frac{t^{1-\mu_1}}{\Gamma(2-\mu_1)}+\alpha_2\frac{t^{1-\mu_2}}{\Gamma(2-\mu_2)},\\ & 1<\mu_1<\mu_2<2 \quad \textrm{or} \quad 0<\mu_1<\mu_2<1\end{array}$       & $\quad \sim t^{\mu_2}$       & $\quad \sim t^{\mu_1}$ \\ \hline
  $\begin{array}{c l} & \sum_{j=1}^{N}\alpha_j\frac{t^{1-\mu_j}}{\Gamma(2-\mu_j)},\\ & 1<\mu_1<\dots<\mu_N<2 \quad \textrm{or}\\ & 0<\mu_1<\dots<\mu_N<1\quad \end{array}$       & $\quad \sim t^{\mu_N}$       & $\quad \sim t^{\mu_1}$ \\ \hline
  $\quad \int_{1}^{2}\frac{t^{1-\lambda}}{\Gamma(2-\lambda)}\,\mathrm{d}\lambda, \quad 1<\lambda<2 \quad$       & $\quad \sim t^2\log{\frac{1}{t}}$       & $\quad \sim t\log{t}$ \\ \hline
  $\quad e^{-bt}\frac{t^{1-\mu}}{\Gamma(2-\mu)}, \quad 1<\mu<2, \quad b>0 \quad$       & $\quad \sim t^\mu$       & $\quad \sim t^2$ \\ \hline
  $\begin{array}{c l} & e^{-bt}t^{\beta-1}E_{\alpha,\beta}^{\delta}\left(-\nu t^{\alpha}\right), \quad 0<\alpha\delta<\beta<1, \\ &\nu>0, \quad b>0 \quad \end{array}$       & $\quad \sim t^{2-\beta}$       & $\quad \sim t^2$ \\ \hline
  $\begin{array}{c l}& t^{1-\mu}E_{\rho,2-\mu}^{-\delta}\left(-\nu t^{\rho}\right), \\ & 1<\mu<2, \quad 0<2-\mu<\delta, \quad \\ & 2-\delta<\mu-\rho\delta<2 \quad 0<\rho,\delta<1\end{array}\quad$       & $\quad \sim t^{\mu}$       & $\quad \sim t^{\mu-\rho\delta}$ \\ \hline
  $\begin{array}{c l}&\int_{1}^{2}t^{1-\mu}E_{\rho,2-\mu}^{-\delta}\left(-\nu t^{\rho}\right), \\ & 1<\mu<2, \quad 0<2-\mu<\delta, \quad \\ & 2-\delta<\mu-\rho\delta<2 \quad 0<\rho,\delta<1\end{array}$       & $\quad \sim t^{2}\log{\frac{1}{t}}$       & $\quad \sim t^{1-\rho\delta}\log{t}$ \\ \hline
  \end{tabular}
\end{table}

It would be of interest to study the case with distributed order memory kernel of the form $\eta(t)=\int_{0}^{2}p(\lambda){_{C}D_{t}^{\lambda}}\,d\lambda$, which is a kind of generalization of the telegrapher's equation \cite{masoliver,masoliver2,orsinger}. 
We also leave for future investigation the corresponding model in the Riemann-Liouville form, which is given by
\begin{equation}
\frac{\partial^2}{\partial t^2}W(x,t)=\frac{\partial^{2}}{\partial t^2}\int_{0}^{t}\zeta(t-t')\frac{\partial^2}{\partial x^2}W(x,t')\,dt'.
\end{equation}

\section*{Acknowledgements:}
{ZT is supported by NWO grant number 040.11.629, Department of Applied Mathematics, TU Delft. TS and AC acknowledge support within DFG (Deutsche Forschungsgemeinschaft) project ``Random search processes, L\'evy flights, and random walks on complex networks", ME 1535/6-1.}

\end{document}